\documentclass[authoryear,11pt]{article}
\usepackage{natbib}
\usepackage[english]{babel}
\usepackage{amsmath}
\usepackage{latexsym}
\usepackage{amssymb}
\usepackage{epsfig}
\usepackage[normalem]{ulem}
\usepackage[usenames,dvipsnames]{xcolor}
\newcommand{\ci}{\!\perp \! \! \! \perp\!}
\newcommand{\nci}{\mbox{\protect $\: \perp \hspace{-2.3ex}\perp $} \hspace{-2mm}/ \hspace{1.5mm}}

\newtheorem{defin}{Definition}

\newtheorem{thm}{Theorem}

\newtheorem{prop}[thm]{Proposition}
\newenvironment{proof}{\begin{trivlist}\item[] \mbox{\textit{Proof.}}}
{\hfill$\Box$ \end{trivlist}}
\begin{document}
%
\title{Conditional and marginal relative risk parameters for a class of recursive  regression graph models}
\author{Monia Lupparelli \\Department of Statistical Sciences, University of Bologna, IT\\\textit{monia.lupparelli@unibo.it}}
\date{}

\maketitle
\begin{abstract}
\noindent In linear  regression modelling the  distortion of effects  after marginalizing over variables of the conditioning set has been widely studied in several contexts. For Gaussian variables, the relationship between marginal and partial regression coefficients is well-established and 
the issue is often addressed as a result of W. G. Cochran. 
 Possible generalizations  beyond the linear Gaussian case have been developed, nevertheless the case of discrete variables is still challenging, in particular in medical and social science settings. A  multivariate regression framework  is proposed for binary data with regression coefficients given by the logarithm of relative risks and a multivariate Relative Risk formula is derived to define the relationship between marginal and conditional relative risks. The method is illustrated through the analysis of the morphine data in order to assess the effect of preoperative oral morphine administration on the postoperative pain relief. 
\end{abstract}

\section{Introduction}
A  regression framework is adopted for modelling the effect of a set of explicative variables  on a set of dependent variables. Explicative variables are sometimes called explanatory variables or predictors as well as dependent variables are also called response variables or outcomes. 
Consider three Gaussian variables: a response variable $Y$ and two explicative variables $\{Z,X\}$. The  linear regression model $E(Y|\{Z,X\})=\beta_{\emptyset}+ \beta_{Y|Z.X}Z+ \beta_{Y|X.Z}X  $ includes the intercept $\beta_{\emptyset}$ and the partial regression coefficients $\beta_{Y|Z.X}$ and $\beta_{Y|X.Z}$, respectively of $Z$ and $X$.  
If the interest is in marginal rather than in conditional associations, the \textit{marginal effect} $\beta_{Y|X}$ obtained marginalizing over $Z$ is expected to be different from the \textit{conditional effect} $\beta_{Y|X.Z}$. 

\begin{figure}[b]
\centering\caption{DAG models: (a) independence model $Y \ci X|Z$; (b)  saturated model.}\label{fig.med}
    \includegraphics[scale=0.7]{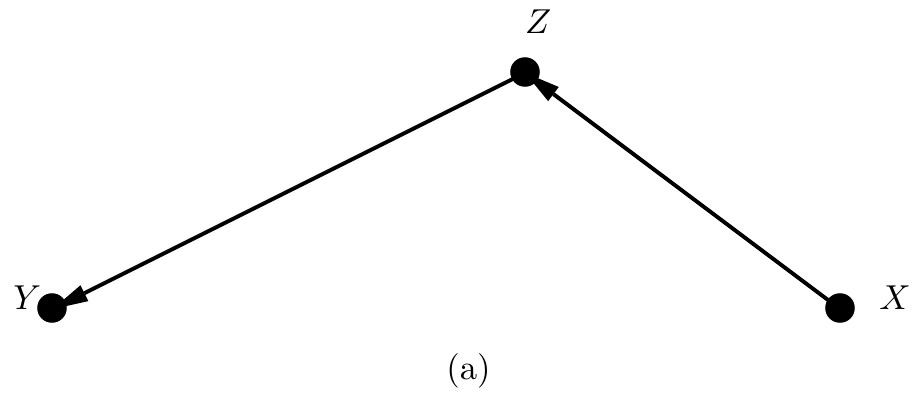}
	\includegraphics[scale=0.7]{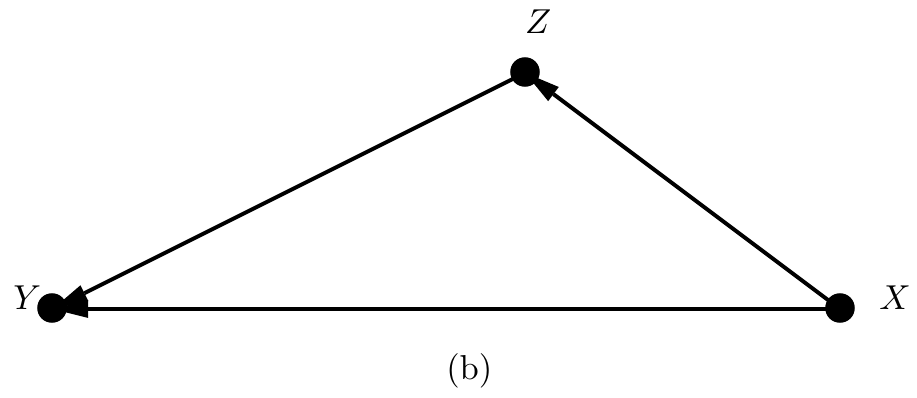}	
\end{figure}

An extreme example  is given by the DAG in Figure \ref{fig.med}(a). 
The statistical model corresponds to the  recursive regression of $Y$ on $Z$ and $X$  and the regression of $Z$ on $X$. Then, $X$ is a pure explanatory variable  for both $Z$ and $Y$ and $Z$ is an \textit{intermediate variable} because it is a response with respect to $X$ and an explanatory with respect to $Y$. Under  Markov properties defined for DAGs, missing arrows imply conditional independencies for  variables associated to pairs of disjoined nodes. The missing arrow between  $Y$ and $X$ means  $Y \ci X |Z$, in case of linear regressions this implies $\beta_{Y|X.Z}=0$ which is expected to be  different from $\beta_{Y|X}$; see \citet{wermuth2012sequencies}. 

For Gaussian variables the linear relationship between  marginal and conditional  regression coefficients is well-established in the context of path analysis.  In particular, \citet{cochran:1938} represents a notable reference  such that this  relationship  is also known as the \textit{Cochran's formula}:
\begin{equation}\label{eq:Cochran}
	\beta_{Y|X}=\beta_{Y|X.Z}+ \beta_{Y|Z.X}\beta_{Z|X}.
\end{equation}
So $\beta_{Y|X}$ is sometimes called the overall or total effect of $X$ on $Y$ obtained linearly combining  the direct effect $\beta_{Y|X.Z}$ with the indirect effect  $\beta_{Y|Z.X}\beta_{Z|X}$ \citep{wer-cox:2015}. This represents a typical framework for mediation analysis where $X$ is a treatment, $Y$ is an outcome and  $Z$ is a mediator of the effect of $X$ on $Y$; see  Figure \eqref{fig.med}(b) for a graph representation and \citet{vanderweele:2016} for a recent review. In  social science settings, when the intermediate variable is discrete,  the latter effect is sometimes known as the moderating effect; see \citet{wermuth:1987}.
Regardless of any context,  this effect will be  denoted as \textit{deviation term}, given that it represents the deviation between the marginal and the conditional effect of $X$ on $Y$.

Possible generalizations of the Cochran's formula have been investigated for non-Gaussian distributions. 
 \citet{cox-wer:1994}  derived a formula for logistic regression models  assuming a quadratic exponential distribution. \citet{wer-et-al:2009}  proved that the formula holds for the special case of palindromic distributions; see  \citet{wer-mar:2018} for more recent results related to palindromic Ising models. \citet{cox:2007} generalized the Cochran's formula  for a non-linear quantile regression approach when all variables are continuous, and extensions to discrete variables are only outlined. 
Further extensions, even if not directly addressed as generalization of the Cochran's formula, have been investigated. \citet{van-van:2010}  proposed a logistic regression approach for mediation analysis  when the mediator  is continuos.  In a similar context, \citet{stan-dor:2018} explored the relation between marginal and conditional parameters in logistic regression models.   
In the context of  confounders, when the   distortion is given by ignoring  an unobservable  background variable, \citet{lin-al:1998} provided  substantial results beyond the Gaussian case.

Exploring a close relationship between marginal and conditional effects in discrete regression models still represents a crucial issue. Furthermore,  the generalization for the multivariate case involving  random vectors $Y_V=(Y_v)_{v \in V}$ and $Z_U=(Z_u)_{u \in U}$ of non-independent outcomes and  intermediate variables, respectively, seems to be  unexplored. 

This paper proposes a framework of multivariate recursive regressions so that a counterpart of the Cochran's formula can be derived for binary variables and generalized for the case of multiple response and intermediate variables. The link function adopted in these regressions is linear in the logarithm of the probabilities and the  coefficients in single regressions are log-relative risks. The interpretation of the  coefficients in terms of relative risks is preserved even for multivariate regressions.  This regression approach represents a special case in the class of log-mean linear regression models of \citet{lup-rov:2017} who  developed the main statistical properties and discussed algorithms for maximum likelihood estimation.

 Then,  a \textit{Relative Risk formula} results: the marginal relative risk of each  outcome $Y_v \in Y_V$ associated with $X$ is obtained combining the conditional relative risk given the intermediate variable $Z_U$  with a deviation term. An interesting interpretation is provided for the deviation term. The simple univariate case is illustrated with the analysis of the smoking habits data aimed to assess the effect of parents and siblings smoking habits on the smoking behavior of college students \citep{Spielberger-al:1983}. The Relative Risk formula is applied to address 
   the moderating effect provided by considering the two distinct sub-groups of teenagers with different siblings smoking habits. 

More interesting is  the multivariate  case involving  multiple outcomes and multiple intermediate variables here discussed through the analysis of the morphine case study.  This is a prospective, randomized double-blind clinical study which aims to assess the effect of preoperative administration of oral morphine on postoperative pain relief observed in two distinct time occasions after the surgery, in order to reduce the use of postoperative morphine; see \citet{Borracci2013}. The data set used in this work involves a randomized treatment, two final outcomes representing the pain intensity at rest and on movement (i.e., upon coughing) observed  24 hours after the surgery and two intermediate variables given by the same pain indicators observed after 4 hours. The static and dynamic pain indicators are useful to explore how the treatment acts on different kinds of pain over the time. An univariate regression approach is not suitable because the pain intensity at rest and on movement  are reasonable assumed to be non-independent both after 4 and 24 hours. Then, the proposed multivariate regression framework is applied for the analysis of the morphine data and the Relative Risk formula is used to estimate the overall effect of oral preoperative  morphine on pain relief and, in particular, to disentangle the direct effect of the treatment on the final pain intensity after 24 hours and the indirect effect through the pain intensity after 4 hours.
 
\section{The regression framework} \label{sec:lm}
\subsection{Relative risk-based measures of association}\label{ss:RR}
We consider a vector $(Y, Z, X)$ of three binary variables taking value $i \in \{0,1\}^3$. In particular, $Y$ is the final outcome equal to 1 if the event of interest occurs, $X$ is a pure  explanatory variable and $Z$ is an intermediate variable. Relevant relative risks for the event $\{Y=1\}$ associated with $Z$ and $X$ are defined.
Let
\begin{equation}\label{eq:conditionalRR-YX|Z}
RR_{Y|X.Z=0}=\frac{P(Y=1|X=1, Z=0)}{P(Y=1|X=0, Z=0)}, \qquad RR_{Y|Z.X=0}=\frac{P(Y=1|Z=1, X=0)}{P(Y=1|Z=0, X=0)}
\end{equation} 
be the \textit{conditional relative risk} of $Y$ associated with $X$, given $Z=0$ 
and the conditional relative risk  of $Y$ associated with $Z$, given $X=0$, respectively. Also, consider the  interaction term
\begin{equation}\label{eq:interactionRR-YXZ}
	RR_{Y|ZX}=\frac{P(Y=1|Z=1, X=1) \times P(Y=1|Z=0, X=0)}{P(Y=1|Z=1, X=0) \times P(Y=1|Z=0, X=1)}.
\end{equation}  
Therefore, conditional relative risks in Equations \eqref{eq:conditionalRR-YX|Z}  for different values of the conditioning set can be easily derived:
\begin{equation}\label{eq:conditionalRR-cs=1}
	RR_{Y|X.Z=1} =RR_{Y|X.Z=0} \times RR_{Y|ZX} \quad \text{and} \quad RR_{Y|Z.X=1} =RR_{Y|Z.X=0} \times RR_{Y|ZX}.
\end{equation}
If the intermediate variable is ignored, let
\begin{equation}\label{eq:marginalRR}
	RR_{Y|X}=\frac{P(Y=1|X=1)}{P(Y=1|X=0)} 
\end{equation}
be the \textit{marginal relative risk}  of the outcome $Y$  associated with the explanatory variable $X$. In general $RR_{Y|X.Z}$ is expecetd to be different from $RR_{Y|X}$. Similarly, considering the event $\{Z=1\}$, 
\begin{equation}\label{eq:marginalRR-ZX}
	RR_{Z|X}=\frac{P(Z=1|X=1)}{P(Z=1|X=0)}
\end{equation}
is the marginal relative risk of the intermediate variable $Z$ associated with the background variable $X$. 
\subsection{The log-mean regression model}
Given the joint distribution $p$ of the random vector $(Y, Z, X)$, consider the factorization
\begin{equation}
	p=p_{Y|ZX}\times p_{Z|X} \times p_X. 
\end{equation}
based on the DAG model in Figure\ref{fig.med}(b). Hereafter, in the subscript the short notation $Y|ZX$  is adopted instead of $Y|Z \cup X$.
The probability function $p_{Y|ZX}$  is a Bernoulli distribution with probability parameter $\pi_{Y|i_{ZX}}$, for any value $i_{ZX} \in \{0,1\}^2$ of the conditioning set. Similarly, $p_{Z|X}$  is a Bernoulli distribution with probability parameter $\pi_{Z|i_{X}}$, with $i_X \in \{0,1\}$.

A log-mean  regression framework is adopted for modelling  via a linear predictor the  logarithm of the probability parameters of the   distributions $p_{Y|ZX}$ and  $p_{Z|X}$. A simplified notation for binary variables is used on the same fashion of the linear regression case, then,
\begin{eqnarray}\label{eq:reg1bis}
\log \pi_{Y|i_{ZX}}&=& \alpha_{Y|ZY}+ \theta_{Y|Z.X}Z+\theta_{Y|X.Z}X+\theta_{Y|ZX}ZX, \qquad i_{ZX} \in \{0,1\}^2,\\\label{eq:reg2}
\log \pi_{Z|i_X} &=& \alpha_{Z|X} + \theta_{Z|X}X, \qquad \qquad \qquad \qquad \qquad \qquad  \quad \;\;\; i_{X} \in \{0,1\}.
\end{eqnarray}
For regression model \eqref{eq:reg1bis}, the intercept is given by $\alpha_{Y|ZX}$. Parameters $\theta_{Y|Z.X}$ and $\theta_{Y|X.Z}$ are the main effect of $Z=1$ and of $X=1$, respectively, on the response variable $Y$, and $\theta_{Y|ZX}$ is the effect of the interaction between $Z$ and $X$. These regression coefficients are the logarithm of the conditional relative risks introduced in Section \ref{ss:RR}:
\begin{equation}
	RR_{Y|Z.X=0}=\exp(\theta_{Y|Z.X} ), \quad   RR_{Y|X.Z=0}=\exp(\theta_{Y|X.Z} ), \quad RR_{Y|ZX}=\exp(\theta_{Y|ZX} ),
\end{equation} 
such that $RR_{Y|Z.X=1}= \exp(\theta_{Y|Z.X}+\theta_{Y|ZX})$ and $RR_{Y|X.Z=1}= \exp(\theta_{Y|X.Z}+\theta_{Y|ZX})$.

For regression model in Equation \eqref{eq:reg2}, the intercept is $\alpha_{Z|X}= \log \pi_{Z|\{X=0\}}$, also
$$
RR_{Z|X}=\exp(\theta_{Z|X}).
$$

The regression of $Y$ on $X$ when removing $Z$ from the conditioning set of $Y$ is given by
\begin{equation}\label{eq:reg-margYX}
	\log \pi_{Y|i_X}= \alpha_{Y|X}+\theta_{Y|X}X, \qquad i_X \in \{0,1\};
\end{equation}
coefficient $\theta_{Y|X}$ is the logarithm of the marginal relative risk in Equation \eqref{eq:marginalRR}. 
\section{The Relative Risk formula}
Combining  regression coefficients in Equations \eqref{eq:reg1bis} and \eqref{eq:reg2}, the 
marginal coefficient when regressing $Y$ on $X$ in Equation \eqref{eq:reg-margYX} can be obtained. 

\begin{prop}\label{thr.main}
Consider the log-mean regression models in Equations \eqref{eq:reg1bis} and \eqref{eq:reg2} for the random binary vector $(Y,Z,X)$. The  marginal log-mean regression coefficient when regressing $Y$ on $X$ is given by
\begin{equation}\label{eq:thr}
	\theta_{Y|X}= \theta_{Y|X.Z} + \lambda,
\end{equation}
where	
$$
\lambda=	\log \frac{\exp(\theta_{Y|Z.X}+ \theta_{Y|ZX})*\exp(\alpha_{Z|X}+\theta_{Z|X})+ 1-\exp(\alpha_{Z|X}+\theta_{Z|X})}{\exp(\theta_{Y|Z.X})*\exp(\alpha_{Z|X})+ 1-\exp(\alpha_{Z|X})}.
$$
\end{prop}
A proof is given in the Appendix.

It can be easily verified that Equation \eqref{eq:thr} can be written in terms of relative risk parameters, then,
\begin{equation}\label{eq:cor}
RR_{Y|X}= RR_{Y|X.Z=0} \times \frac{RR_{Y|Z.X=1} \times  \pi_{Z|\{X=1\}} + (1-\pi_{Z|\{X=1\}})}{RR_{Y|Z.X=0}\times \pi_{Z|\{X=0\}} + (1-\pi_{Z|\{X=0\}})}.	
\end{equation}
See \citet{lin-al:1998} for a close result when the distortion of effects derives from an unmeasured background variable, so that  the conditional probability $\pi_{Z|X}$ is not modelled  in a regression framework and the result depends on the probabilistic assumptions on the unobserved variable.
 
The second factor term in Equation \eqref{eq:cor} corresponds to $\exp(\lambda)$ which represents the so-called deviation term between the marginal and the conditional relative risk of $Y$ associated with  $X$. An interpretation of this term is provided based on the following assumption:
without loss of generality,  let $RR_{Y|Z.X=0}= RR_{Y|Z.X=1}=1$ when $Z=0$.  This implies that $\theta_{Y|Z.X}=\theta_{Y|ZX}=0$ if $Z=0$.

Therefore, the \textit{Relative Risk formula} derives:
\begin{equation}\label{eq:RRF}
		RR_{Y|X}= RR_{Y|X.Z=0} \times \frac{\overline{RR}_{Y|Z.X=1}}{\overline{RR}_{Y|Z.X=0}},
\end{equation}
where $\overline{RR}_{Y|Z.X=1}$ is a weighted average of the conditional relative risk of $Y$ associated with $Z$, given $X=1$, in case $Z=1$ and $Z=0$; as weights the conditional probabilities $P(Z=1|X=1)$ and $P(Z=0|X=1)$ are used, respectively. Similarly $\overline{RR}_{Y|Z.X=0}$ is a weighted average of the conditional relative risk of $Y$ associated with $Z$, given $X=0$, in case $Z=1$ and $Z=0$, with weights given by the conditional probabilities $P(Z=1|X=0)$ and $P(Z=0|X=0)$. 
\begin{figure}[t]\centering\caption{DAG models: (a) independence model $Y \ci Z|X$; (b) independence model $Z \ci X$.}\label{fig.med1}
    \includegraphics[scale=0.7]{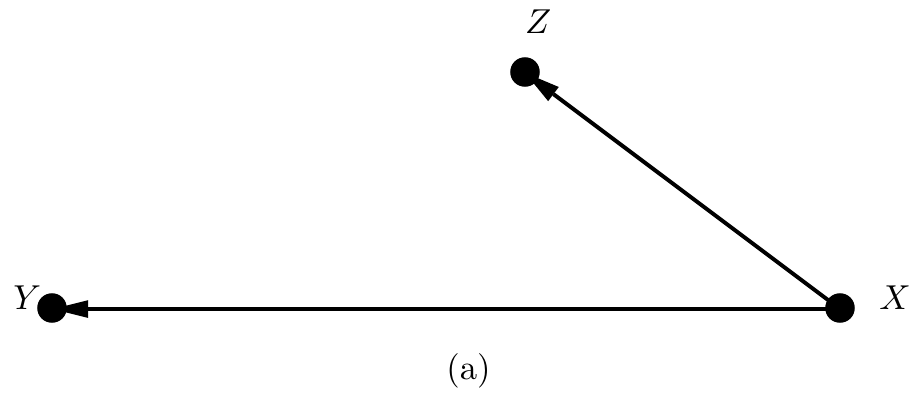}
	\includegraphics[scale=0.7]{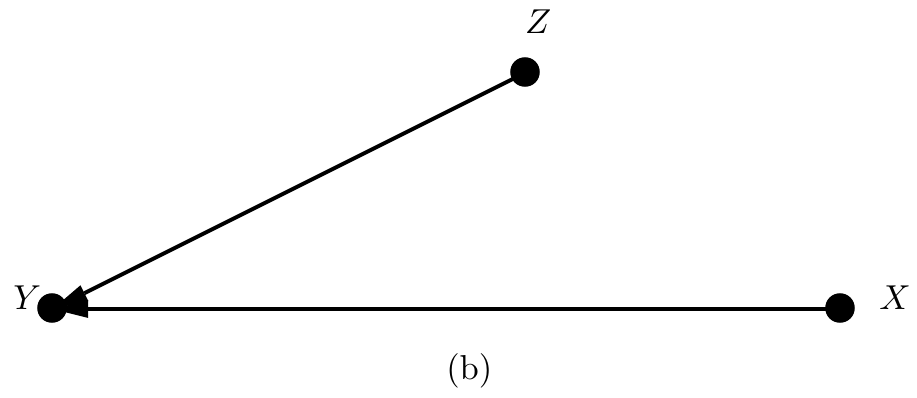}
\end{figure}

The Relative Risk formula closely recalls the Cochran's one. However, there are expected differences given by the different nature of the variables. 
The main focus of the comparison is on the deviation term:  $\exp(\lambda)$  versus $\beta_{Y|Z.X}\beta_{Z|X}$. 

 Firstly, consider the  independence model $Y \ci Z| X$ in Figure \ref{fig.med1}(a) which implies that $\theta_{Y|Z.X}=\theta_{Y|ZX}=0$ \citep[see][]{lup-rov:2017}. Then,  $\lambda=0$ from Proposition \ref{thr.main}, and $RR_{Y|X}=RR_{Y|X.Z}$. For the Gaussian case $\beta_{Y|X}=\beta_{Y|X.Z}$, so in both cases the deviation term is null.
 
Consider the DAG in Figure \ref{fig.med}(a) where $Y \ci X|Z$ implies that $\theta_{Y|X.Z}=\theta_{Y|ZX}=0$, 
then $RR_{Y|X}=\exp(\lambda)$. In linear regressions the independence  model implies $\beta_{Y|X}=\beta_{Y|Z.X}\beta_{Z|X}$. However, notice that given $\theta_{Y|ZX}=0$, $RR_{Y|Z.X=0}=RR_{Y|Z.X=1}$, then $\lambda$ is the log-ratio of the average of the same relative risk values but weighted with different weights.

Finally, consider the  independence model $X \ci Z$ in Figure \ref{fig.med1}(b). For Gaussian variables, the independence implies $\beta_{Z|X}=0$, then, $\beta_{Y|X}=\beta_{Y|X.Z}$. Instead, for the binary case  $\theta_{Z|X}=0$ is not a sufficient condition to have $\lambda=0$, the   constraint only implies that  same probability weights are used in the numerator and in the denominator of the deviation term. Nevertheless, if a model with null interaction term is assumed, i.e,  $\theta_{Y|ZX}=0$,  $\lambda=0$ and $RR_{Y|X}=RR_{Y|X.Z}$.
\begin{table}[t]
\centering\caption{Smoking habits data set.}\label{tab:data}
	\begin{tabular}{ccccc}
\hline\hline
  & \multicolumn{4}{c}{$X$} \\\cline{2-5}
& \multicolumn{2}{c}{$X=0$} & \multicolumn{2}{c}{$X=1$}	\\\cline{2-5}
$Y$ &\multicolumn{1}{c}{$Z=0$} & \multicolumn{1}{c}{$Z=1$}& \multicolumn{1}{c}{$Z=0$} & \multicolumn{1}{c}{$Z=1$}\\\hline
$Y=0$ & 221 & 152 &202  & 196\\
$Y=1$ & 109 & 186 & 158 & 455\\	
\hline\hline
\end{tabular}
\end{table}
\section{An illustrative example: the Smoking habits data}
Consider the set of data taken from \citet{Spielberger-al:1983}  aimed to study  the relation between the family smoking habits and the smoking behaviour of college students. Three binary variables are observed on a sample of 1679 teenager college students: the final response $Y$ which is equal to 1 if the teenager is a smoker and 0 for a non-smoker; the background variable $X$ taking level 1 if both parents are smokers and level 0 if just one of them is a smoker; then, the intermediate variable $Z$ which takes level 1 if siblings are smokers and 0 otherwise. Data are collected in Table \ref{tab:data}. 

It is reasonable to assume that the effect of the parents habits on the teenager smoking behaviour is different within the sub-group of teenagers whose siblings are smokers and the sub-group of teenagers whose siblings are non-smokers. Then, the moderating effect given by the intermediate variable $Z$ needs to be addressed in order to derive the overall effect of the parents habits on the teenager behaviour.
\begin{table}[t]
\centering\caption{Maximum likelihood estimates of log-mean regression coefficients for Smoking habits data. The left block includes estimates and the standard errors of the saturated model ($BIC=6695.879$); with * is denoted a non-significant parameter. The right block includes estimates and the standard errors of the  model with  $\theta_{Y|ZX}=0$ ($BIC=6688.613$). }\label{tab.est}
	\begin{tabular}{lcc|cc}
\hline\hline
Parameters  & Estimates & s.e.  & Estimates & s.e. \\\hline
$\alpha_{Y|ZX}$& -1.108 & 0.078 & -1.086 & 0.056  	\\
$\theta_{Y|Z.X}$& 0.510  & 0.092 & 0.480 & 0.053\\
$\theta_{Y|X.Z}$ & 0.284 & 0.098 & 0.250 & 0.049\\
$\theta_{Y|ZX}$ & -0.045* & 0.113 & - &- \\\hline
$\alpha_{Z|X}$ & -0.681 & 0.038 & -0.681 & 0.038\\	
$\theta_{Z|X}$ & 0.241  & 0.045 & 0.241  & 0.045\\
\hline\hline
\end{tabular}
\end{table}

The regression framework in Equation \eqref{eq:reg1bis} and \eqref{eq:reg2} is fitted for the data representing the DAG model in Figure \ref{fig.med}(b).  Maximum likelihood estimates  and the corresponding standard errors of the regression parameters in the saturated model are collected in Table \ref{tab.est}. The interaction term is shown to be non-significant, then the reduced model  including the constraint $\theta_{Y|ZX}=0$ is fitted providing a deviance 0.16, with 1 degree of freedom, $p$-value=0.69 and a lower BIC value compared to the saturated model; see the estimates in Table \ref{tab.est}.  Then,  the Relative Risk formula is applied in order to derive the estimate of the moderating effect provided by considering the two distinct sub-groups of teenagers with different siblings smoking habits and the estimate of the marginal relative risk between parents and student habits:
\begin{eqnarray*}
	\hat{RR}_{Y|X}&=&\hat{RR}_{Y|X.Z}\times \exp(\hat{\lambda})=\\
	&=&\exp(0.250) \times \frac{\exp(0.480)\times \exp(0.440) + [1-\exp(0.440)]}{\exp(0.480)\times\exp(-0.681) +[1-\exp(-0.681)]}\\
	&=&1.284 \times 1.492=1.914.
\end{eqnarray*} 
\section{The multivariate Relative Risk formula}\label{sec.mrrf}
\begin{figure}[t]
\centering	\caption{Bivariate regression graph models: (a) saturated model; (b) independence model $Y_1 \ci Z|X$, $Y_2 \ci X|Z.$}\label{fig.chain}
\includegraphics[scale=0.6]{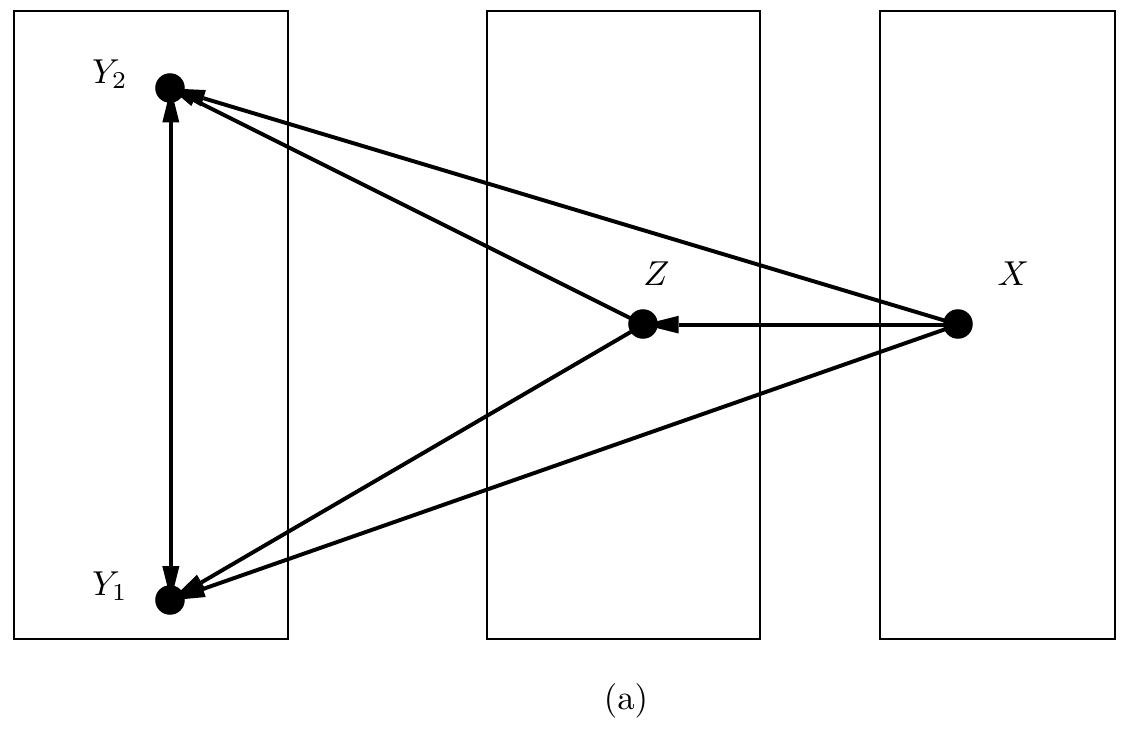} \hspace{1cm}
	\includegraphics[scale=0.6]{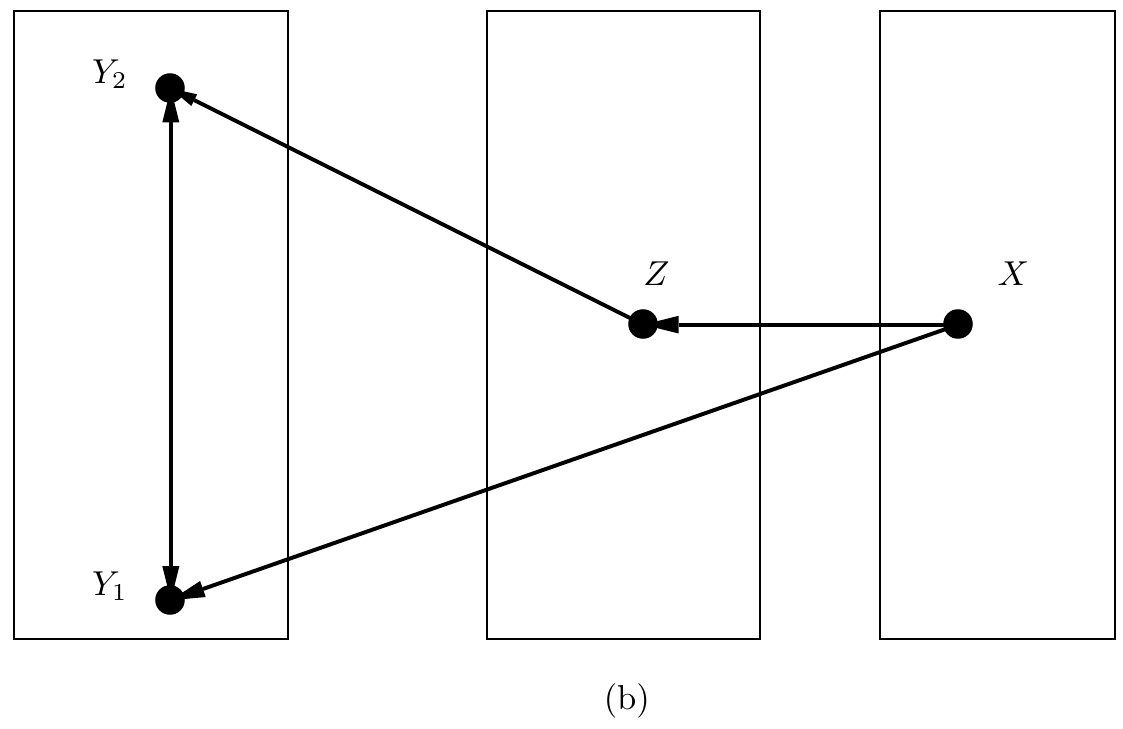}
	\end{figure}

Let $Y_V=(Y_v)_{v \in V}$ be a random vector of binary response  variables. Then, consider the multivariate regression of $Y_V$ on $\{Z,X\}$ which can be represented by   the class of regression graph models \citep{wermuth2012sequencies}. The saturated model for the case of a bivariate vector $Y_V=(Y_1,Y_2)$ is shown in Figure \ref{fig.chain}(a);  variables are partitioned in blocks, variables in different blocks are joined by directed edges preserving the same direction and the response variables collected in the final  block are joined by bi-directed edges denoting that $Y_1 \nci Y_2|\{Z,X\}$. 
In order to derive a multivariate Relative Risk formula, a multivariate log-mean regression framework  is adopted. 

For every subset $D$ of $V$, let $Y_D$ be a marginal vector with marginal probability $\pi_D=P(Y_D=1_D)$, where $1_D$ denotes a vector of $1s$ of length $|D|$. Moreover, consider the conditional probabilities $\pi_{D|i_{ZX}}=P(Y_D=1_D|	\{Z,X\}=i_{ZX})$ and    $\pi_{D|i_X}=P(Y_D=1|X=i_X)$  , with $i_{ZX} \in \{0,1\}^2$, $i_X \in \{0,1\}$ for any $D \subseteq V$. 

The multivariate recursive regression framework for modelling the regression of $Y_V$ on $\{Z,X\}$ and the regression of $Z$ on $X$ is given by 
\begin{eqnarray}\label{eq:mult1}
	\log \pi_{D|i_{ZX}} &=& \alpha_{D|ZX}+ \theta_{D|Z.X}Z + \theta_{D|X.Z}X + \theta_{D|ZX}ZX, \quad i_{ZX} \in \{0,1\}^2, D \subseteq V,\\\label{eq:mult1bis}
	\log \pi_{Z|i_X}&=& \alpha_{Z|X}+ \theta_{Z|X}X, \qquad i_{X} \in \{0,1\}.
\end{eqnarray}
Notice that the model in Equation \eqref{eq:mult1bis} coincides with the model in Equation \eqref{eq:reg2}.

Equation \eqref{eq:mult1} represents a sequence of single and joint regressions for modelling che conditional distribution of $Y_V|\{Z,X\}$; see \citet{lup-rov:2017}. For single regressions of $Y_v$ on $\{Z,X\}$, 
\begin{equation}
	\exp(\theta_{v|Z.X})= RR_{v|Z.X=0} \;\;\; \text{and} \;\;\; \exp(\theta_{v|X.Z})= RR_{v|X.Z=0}, \qquad v \in V;
\end{equation}
the interaction term $\theta_{v|ZX}$ is used to derive conditional relative risks for level 1 of the conditioning variable as in Equation \eqref{eq:conditionalRR-cs=1}. Before to discuss joint regressions, the notion of \textit{product outcome} is introduced. For any non-empty subset $D$ of $V$, let
\begin{equation}
	Y^D=\prod_{v \in D}Y_v
\end{equation} 
be a product outcome, which is a binary variable taking level 1 in case   $Y_D=1_{D}$, and level 0 otherwise. Then, the event $\{Y^D=1\}$ denotes  the co-occurrence of a non-empty subset $D$ of  outcomes. For joint regressions in Equation \eqref{eq:mult1},
\begin{equation}
	\exp(\theta_{D|Z.X})= RR_{D|Z.X=0}  \;\;\; \text{and} \;\;\;  \exp(\theta_{D|X.Z})= RR_{D|X.Z=0}, \qquad D \subseteq V,
\end{equation}
where 
\begin{equation}
	RR_{D|Z.X=0}=\frac{P(Y^D=1|Z=1,X=0)}{P(Y^D=1|Z=0,X=0)},\; RR_{D|X.Z=0}=\frac{P(Y^D=1|X=1,Z=0)}{P(Y^D=1|X=0,Z=0)}
\end{equation}
are the conditional relative risks for the event $\{Y^D=1\}$ associated with $Z$ and $X$, respectively. 
The interaction term $\theta_{D|ZX}$, for any $D \subseteq V$, is used to derive conditional relative risks  for the level 1 of the conditioning variable, as in Equation \eqref{eq:conditionalRR-cs=1}.

If the intermediate variable is ignored, the multivariate log-mean regression of $Y_V$ on $X$ is given by the sequence of regressions
\begin{equation}\label{eq:mult2}
	\log \pi_{D|i_{X}} = \alpha_{D|X} + \theta_{D|X}X, \qquad \qquad i_{X} \in \{0,1\},\;\; D \subseteq V.
\end{equation}
For any $D \subseteq V$, Equation \eqref{eq:mult2} models the conditional distribution of $Y_D|X$; in particular, $\alpha_{D|X}=\log \pi_{D|X=0}$ and $\exp(\theta_{D|X})=RR_{D|X}$ where
\begin{equation}\label{eq:thetaY_V-X}
	RR_{D|X}=\frac{P(Y^D=1|X=1)}{P(Y^D=1|X=0)}, \qquad D \subseteq V
\end{equation} 
is the marginal relative risk of each product outcome $Y^D$ associated with $X$.

Exploiting the properties of the class of log-mean regression models, Proposition \ref{thr.main} can be generalized for the multivariate case.
\begin{prop}\label{thr.mult}
Consider the multivariate log-mean regression models  in Equations \eqref{eq:mult1} and  \eqref{eq:mult1bis} for the random binary vectors $Y_V=(Y_v)_{v \in V}$ and $(Z,X)$. The  marginal log-mean regression coefficients when regressing $Y_V$ on $X$ are given by
	\begin{equation}
	\theta_{D|X}= \theta_{D|X.Z} + \lambda_D, \qquad D \subseteq V,
\end{equation}
where	
$$
\lambda_D=	\log \frac{\exp(\theta_{D|Z.X}+ \theta_{D|ZX})*\exp(\alpha_{Z|X}+\theta_{Z|X})+ 1-\exp(\alpha_{Z|X}+\theta_{Z|X})}{\exp(\theta_{D|Z.X})*\exp(\alpha_{Z|X})+ 1-\exp(\alpha_{Z|X})}.
$$
\end{prop}
A proof is given in the Appendix.

The deviation term $\lambda_D$   preserves the same interpretation given in Equation \eqref{eq:cor} in terms of relative risks with respect to the the event $\{Y^D=1\}$:
\begin{equation}
	\exp(\lambda_D)= \frac{RR_{D|Z.X=1} \times  \pi_{Z|\{X=1\}} + (1-\pi_{Z|\{X=1\}})}{RR_{D|Z.X=0}\times \pi_{Z|\{X=0\}} + (1-\pi_{Z|\{X=0\}})}, \quad D \subseteq V.
\end{equation} 
Therefore, the \textit{Multivariate Relative Risk formula} derives:
\begin{equation}\label{eq:MRR}
	RR_{D|X}= RR_{D|X.Z=0} \times \frac{\overline{RR}_{D|Z.X=1}}{\overline{RR}_{D|Z.X=0}}, \qquad D \subseteq V.
\end{equation}

The multivariate relative risk formula under special  independence assumptions may provide a different decomposition of effects  for each product outcome. For instance consider the regression graph model in Figure \ref{fig.chain}(b), where two missing directed edges 
imply $Y_1 \ci Z|X$ and $Y_2 \ci X|Z$. Applying the formula  for $D=1$, the deviation term $\lambda_1$ vanishes given that $\theta_{1|Z.X}=\theta_{1|ZX}=0$, therefore $RR_{1|X}=RR_{1|X.Z}$. On the other hand, for $D=2$, $RR_{2|X}\neq RR_{2|X.Z}$ because $\lambda_2 \neq 0$ even though the independence constraints $\theta_{2|X.Z}=\theta_{2|ZX=0}=0$.

Interestingly, no simplifications of the formula result with respect to the product outcome $Y^{12}$, because no zero restrictions are implied by the independence statements for the joint regression of $Y_{12}$ on $\{Z,X\}$. Nevertheless, further non-independence constraints might  be included, for instance, if the  probability $\pi_{12|i_{ZX}}$ is invariant given any level $i_{ZX} \in \{0,1\}^2$, then $\theta_{12|Z.X}=\theta_{12|X.Z}=\theta_{12|ZX}=0$,  then the deviation term $\lambda_{12}$ is null and $RR_{12|X}=RR_{12|X.Z}$.
\section{Multiple intermediate variables}\label{sec:intermediate}
\begin{figure}[t]
\centering	\caption{Regression graph models with bivariate outcomes and bivariate intermediate variables : (a) saturated model; (b) independence model $\{Y_R,Y_M\} \ci Z_M|\{Z_R,X\}$}\label{fig.chain-mult}
\includegraphics[scale=0.6]{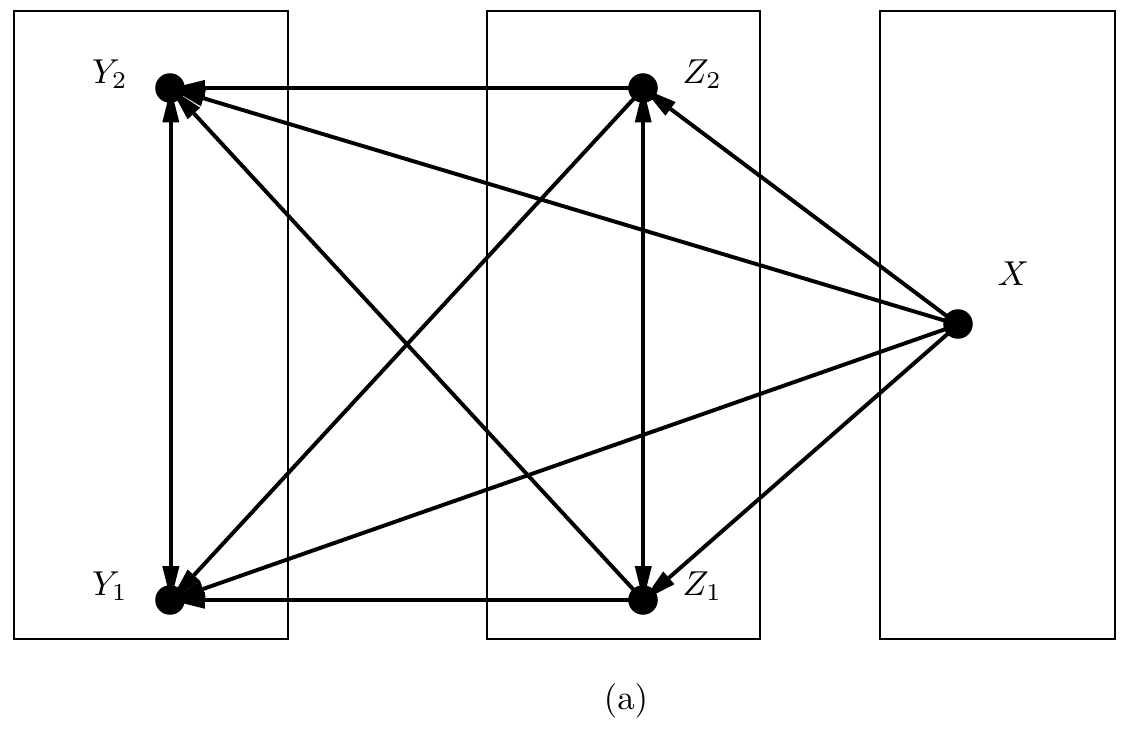} \hspace{1cm}
	\includegraphics[scale=0.6]{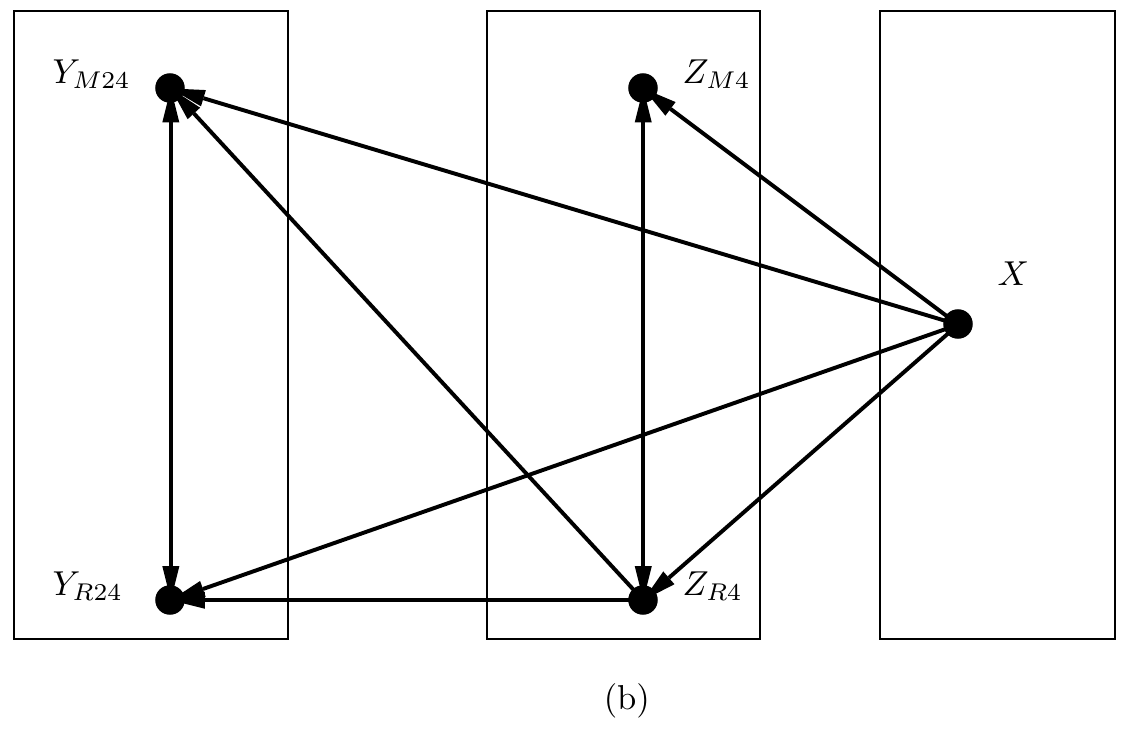}
	\end{figure}

It is also interesting the generalization of the multivariate regression model in Section \ref{sec.mrrf} including a multiple set $Z_U=(Z_u)_{u \in U}$ of non-independent intermediate variables.  For any $D \subseteq V$, $Y^D|\{Z_U,X\}$ is a Bernoulli distribution with probability parameter $\pi_{D|\{i_U,i_X\}}=P(Y^D=1|Z_U=i_U,X=i_X)$ where $i_U \in \mathcal{I}_U=\{0,1\}^{|U|}$ and $i_X \in \{0,1\}$. For any $E \subseteq U$, $Z^E|X$ is a Bernoulli distribution with probability parameter $\pi_{E|i_X}=P(Z^E=1|X=i_X)$, where $Z^E=\prod_{u \in E}Z_u$ is the product intermediate variable, for any $E \subseteq U$. So, $Z^E$ is a binary variable taking level 1 of $Z_E=1_E$, and 0 otherwise, for any $E \subseteq U$. An example in case of bivariate outcomes and bivariate intermediate variables is given in Figure \ref{fig.chain-mult}(a).

The recursive  regression framework is implemented for modelling  the regression of $Y_V$ on $\{Z_U,X\}$ and the regression of $Z_U$ on X:
\begin{eqnarray}\label{eq:regmultVU}
\log \pi_{D|\{i_{U},i_X\}}&=& \alpha_{D|UX}+ \sum_{u \subseteq U}\theta_{D|u.X}Z_u +\theta_{D|X.U}X,  \qquad D \subseteq V \\\label{eq:regmultUX}
\log \pi_{E|i_X} &=& \alpha_{E|X} + \theta_{E|X}X, \qquad\qquad \quad  \qquad \qquad \; E \subseteq U,
\end{eqnarray}
with $i_X \in \{0,1\}$ and $i_U \in \mathcal{I}_U$. Parameters $\theta_{D|u.X}$ and $\theta_{D|X.U}$ are the main effect of $Z_u=1$, for each $u \in U$, and of $X=1$, respectively, on the response variable $Y^D$.   These regression coefficients are the logarithm of the conditional relative risks introduced in Section \ref{ss:RR}:
\begin{equation}\label{eq:thetaY_V-Z}
	RR_{D|u.X=0}=\exp(\theta_{D|u.X} )= \frac{P(Y^D=1|Z_u=1,Z_{U \setminus u}=0_{U \setminus u},X=0)}{P(Y^D=1|Z_U=0_U,X=0)} \quad u \in U, \; D \subseteq V,
\end{equation}  
and
\begin{equation}
	   RR_{D|X.U=0_U}=\exp(\theta_{Y|X.U} )= \frac{P(Y^D=1|Z_U=0_U,X=1)}{P(Y^D=1|Z_U=0_U,X=0)} , \quad D \subseteq V.
\end{equation}  
The model in Equation \eqref{eq:regmultVU} does not include  the interaction terms  among the intermediate variables $Z_E$, with $E \subseteq U$, and  the background variable $X$. Without loss of generality, this simplified model is  assumed in order to make the resulting Relative Risk formula more interpretable. Under this assumption, specifying the level of the conditioning set in the relative risk notation is not required, given that $RR_{D|u.X=1}=RR_{D|u.X=0}$ and $RR_{D|X.U=i_U}=RR_{D|X.U=i_{U'}}$, for any $D \subseteq V$, $u \in U$ and any $i_U, i_{U'} \in \mathcal{I}_U$. So, in the sequel, the shorthand notation is used, e.g, $RR_{D|u.X}$ jointly for $RR_{D|u.X=0}$ and $RR_{D|u.X=1}$ and $RR_{D|X.U}$  instead of $RR_{D|X.U=i_U}$ for any $i_U \in \mathcal{I}_U$.

From Equation \eqref{eq:regmultUX}, the regression of $Z_U$ on $X$ is then modelled via a sequence of regressions of any $Z^E$ on $X$ where  
\begin{equation}\label{eq:thetaZ_U}
	RR_{E|X}=\exp(\theta_{E|X})= \frac{P(Z^E=1|X=1)}{P(Z^E=1|X=0)}, \qquad E \subseteq U,
\end{equation}
is the relative risk for the event $\{Z^E=1\}$ associated with $X$.

The marginal model obtained ignoring all the intermediate variables is given by
\begin{equation}\label{eq:regmultV}
	\log \pi_{D|i_X} = \alpha_{D|X} + \theta_{D|X}X, \qquad  D \subseteq V,
\end{equation}
with $i_X \in \{0,1\}$ and with $\theta_{D|X}$ defined in Equation \eqref{eq:thetaY_V-X}. Then, the following theorem generalizes Proposition \ref{thr.mult} in order to define the close relationship between marginal and conditional relative risk parameters for a class of recursive regression models including both multiple response variables and multiple intermediate variables. 

\begin{thm}\label{thr.multVU}
Consider the multivariate log-mean regression models  in Equations \eqref{eq:regmultVU} and  \eqref{eq:regmultUX} for the random binary vectors $Y_V=(Y_v)_{v \in V}$, $Z_U=(Z_u)_{u \in U}$ and $X$. The  marginal log-mean regression coefficients when regressing $Y_V$ on $X$ are given by
	\begin{equation}
	\theta_{D|X}= \theta_{D|X.U} + \lambda_D, \qquad D \subseteq V,
\end{equation}
where	
$$
\lambda_D= \log \frac{\sum_{E \subseteq U}\exp \Bigg [\sum_{u \in  E}\theta_{D|u.X}Z_{u} \Bigg ] P(Z_E=1_E, Z_{U \setminus E}=0_{U \setminus E}|X=1)}{\sum_{E  \subseteq U}\exp \Bigg [\sum_{u \in  E}\theta_{D|u.X}Z_{u}\Bigg ] P(Z_E=1_E, Z_{U \setminus E}=0_{U \setminus E}|X=0)}, \quad D \subseteq V.
$$
\end{thm}
See the Appendix for the proof. 

The deviation term still represents a ratio of a weighted average relative risk in case $X=1$ and $X=0$ such that the multivariate Relative Risk formula for multiple intermediate variables derives. Then,

\begin{equation}\label{eq:RRYU}
	RR_{D|X}=RR_{D|X.U}\times \frac{\overline{RR}_{D|U.X=1}}{\overline{RR}_{D|U.X=0}}
\end{equation}
where
$$
\overline{RR}_{D|U.X=1}= \sum_{E \subseteq U} RR_{D|E.X}P(Z_E=1_E,Z_{U \setminus E}=0_{U\setminus E}|X=1)
$$
and
$RR_{D|E.X} =\prod_{u \in E} RR_{D|u.X}$ given that a model with no interaction terms is assumed. $\overline{RR}_{D|U.X=0}$ is derived accordingly using different probability weights $P(Z_E=1_E,Z_{U \setminus E}=0_{U\setminus E}|X=0)$ in case of no treatment assignment, for any $E \subseteq U$.

The Relative Risk formula in Equation \eqref{eq:RRYU} can be also derived  for a subset of intermediate variables.
\section{The Morphine study}
The morphine study is a prospective, randomized, double-blind study which aims to investigate the effect of preoperative oral administration of morphine sulphate on postoperative pain relief in order to reduce the postoperative administration of IntraVenous Patient Controlled Analgesia; see \citet{Borracci2013}. A sample of 60 patients is considered, aged between 18-80 and  undergoing and elective open colorectal abdominal surgery. Before surgery, 32 patients were randomly assigned to the treatment group, $X=1$, receiving oral morphine sulphate (Oramorph$^{\textregistered}$, Molteni Farmaceutici, Italy) and 28 patients to the control group, $X=0$, receiving oral midazolam (Hypnovel$^{\textregistered}$, Roche, Switzerland), considered as an active placebo.

The outcomes of interest are the postoperative pain intensity measured through a visual analogue scale at rest and for movement (e.g., upon coughing), in particular measured  4 and 24 hours after the end of the surgery. Visual analogue scale scores are measured using a 100 mm line where no pain and extreme pain are respectively given by the left and the right extremities. Based on physician considerations, 30 mm and 45 mm are considered as cut points for the pain score at rest and on movement, respectively, in order to define a satisfactory postoperative pain relief; see \citet{Borracci2013}.
 
Then, four binary variables result: $Y_{R24}$ and $Y_{M24}$ denoting the pain intensity after 24 hours at rest and on movement, respectively; $Z_{R4}$ and $Z_{M4}$ denoting the pain intensity after 4 hours at rest and on movement, respectively. These binary variables take level 1 in case of a satisfactory pain relief and level 0 otherwise. Then, the variables $Y_{R24}$ and $Y_{M24}$ represent the final outcomes of interest, $Z_{R4}$ and $Z_{M4}$ represent the intermediate outcomes. The product outcome $Y^{\{R24,M24\}}$ and the product intermediate variable $Z^{\{R4,M4\}}$ are also considered; they represent the joint static and dynamic pain intensity after 24 and 4 hours, respectively, so that level 1 corresponds to a satisfactory  pain level both at rest and on movement, at each occasion. 

Reasonable assumptions are that both the static and the dynamic pain intensity are not independent at each time occasion, and that the pain intensity 24 hours after the surgery depends on the pain intensity after 4 hours. The aim of the analysis is to explore how the treatment acts on different kinds of pain over the time in order to reduce the use of postoperative morphine. In particular, the interest is twofold:  (i) assessing the overall treatment effect on the pain relief after 24 hours   and (ii) distinguishing between the direct effect that the treatment still has in reducing (or not reducing)  the  pain intensity after 24 hours and the indirect effect given by the reduction (or not reduction) the treatment acts on the pain level after 4 hours. 

The log-mean regression framework illustrated in Section \ref{sec:intermediate}  is fitted for the complete graph in Figure \ref{fig.chain-mult}(a) and a good statistical fitting results: the deviance is 14.77, with 12 degree of freedom and $p$-value=0.25 ($BIC$=279.98). After a back forward stepwise selection  procedure, the more parsimonious model represented in Figure \ref{fig.chain-mult}(b) has been chosen. The deviance is 18.88, with 15 degree of freedom and $p$-value=0.22 ($BIC$=271.81). The selected model implies $\{Y_{R24},Y_{M24}\} \ci Z_{M4}|\{Z_{R4}, X\}$, that is, both types of pain intensity at 24 hours do not depend on the dynamic pain intensity at 4 hours given the static pain intensity at 4 hours and the treatment assignment, showing that the static pain  represents the crucial indicator  for postoperative pain relief. Moreover, the  model supports the hypothesis that an univariate regression approach would be not appropriate as both  pain indicators are not independent under the selected model.
\begin{table}[h]
\centering\caption{Maximum likelihood estimates and standard errors of log-mean regression coefficients for morphine data, under the regression graph model in Figure \ref{fig.chain-mult}(b). The left-side block includes the estimates of the single regressions and the right-side one the estimates of the joint regressions.}\label{tab.mor}
	\begin{tabular}{lcc|lcc}
\hline\hline
Parameters                  & Estimates & s.e. & Parameters                  & Estimates & s.e.   \\\hline
$\alpha_{R24|\{R4,M4,X\}}$  & -1.040    & 0.234  & $\alpha_{\{R24,M24\}|\{R4,X\}}$  & -2.332    & 0.492    \\
$\theta_{R24|R4}$           & 0.630     & 0.248 & $\theta_{\{R24,M24\}|R4}$  & 0.692    & 0.442 \\
$\theta_{R24|X}$            & 0.329    & 0.217  & $\theta_{\{R24,M24\}|X}$  & 1.187    & 0.496  \\\hline
$\alpha_{M24|\{R4,X\}}$  & -2.055 & 0.416    \\
$\theta_{M24|R4}$           & 0.514     & 0.364 \\
$\theta_{M24|X}$            & 1.096     & 0.449  \\\hline\hline
$\alpha_{R4|X}$            & -1.366    & 0.309  & $\alpha_{\{R4,M4\}|X}$  & -2.511    & 0.575\\
$\theta_{R4|X}$           & 1.060     & 0.324 & $\theta_{\{R4,M4\}|X}$  & 1.731    & 0.602 \\\hline
$\alpha_{M4|X}$            & -2.024    & 0.432\\
$\theta_{M4|X}$           & 1.285     & 0.466 \\\hline\hline

\end{tabular}
\end{table}

Parameter estimates in Table \ref{tab.mor} show a positive effect of the treatment in reducing the pain intensity at rest, on movement and jointly at rest and on movement, at each time occasion. In particular, the estimates of the conditional relative risks for the pain intensity after 24 hours associated with the treatment (given the  static pain intensity after 4 hours) are
\begin{equation}\label{eq:morp1}
	\hat{RR}_{R24|X.R4}=1.390, \quad \hat{RR}_{M24|X.R4}=2.992, \quad \hat{RR}_{\{R24,M24\}|X.R4}=3.277,
\end{equation}
and the estimates of the relative risks for the pain intensity after 4 hours associated with the treatment are
\begin{equation}\label{eq:morp2}
\hat{RR}_{R4|X}=2.887, \quad \hat{RR}_{M4|X}=3.615, \quad \hat{RR}_{\{R4,M4\}|X}=5.646.
\end{equation}
So the preoperative oral morphine has a strong effect in  reducing  both static and dynamic pain; in particular, it is more effective in reducing the dynamic pain (especially after 4 hours) rather than the static one. 

Furthermore, also the pain relief at rest after 4 hours positively influences  both the final pain intensity, and the estimates of the corresponding conditional relative risks  (given the treatment assignment) are
\begin{equation}\label{eq:morp3}
	\hat{RR}_{R24|R4.X}=1.878, \quad \hat{RR}_{M24|R4.X}=1.672, \quad \hat{RR}_{\{R24,M24\}|R4.X}=1.998.
\end{equation}

Estimates in Equation \eqref{eq:morp2} represents the direct effect of the preoperative oral morphine of the final pain relief. Combining estimates in Equations \eqref{eq:morp2} and \eqref{eq:morp3} following the result of Theorem \ref{thr.multVU}, the estimates of the indirect effect of treatment  can be derived, and, consequently, the estimate of the overall (marginal) effect of treatment on the final pain relief at rest, on movement and jointly at rest and on movement. Then, the estimates of the deviation terms are 
\begin{equation}
	\exp(\hat{\lambda}_{R24})=1.345, \qquad \exp(\hat{\lambda}_{M24})=1.276, \qquad \exp(\hat{\lambda}_{\{R24,M24\}})=1.383. 
\end{equation}
The estimates of the marginal treatment effect are
\begin{equation}
	\hat{RR}_{R24.X}=1.390 \times 1.345=1.870, \qquad \hat{RR}_{M24.X}=2.992 \times 1.276=3.818, 
\end{equation}
and
\begin{equation}
	 \hat{RR}_{\{R24,M24\}.X}=3.277 \times 1.383=4.532.
\end{equation}

Concluding, the preoperative morphine has an overall strong effect in reducing postoperative pain intensity after 24 hours. These effects are obtained combining the conditional relative risk and the deviation term of each outcome and product outcome. These two effects are comparable for the after 24 hours static pain intensity, instead the direct effect is stronger than the indirect one in the remaining cases. In particular the direct effect of the treatment is much stronger in improving jointly  the final pain relief at rest and on movement. 
\section{Discussion}
Regression frameworks based on further link functions could be explored in order to derive similar formulas  for different measures of association, such as the odds ratio in logistic regressions, 
however the resulting  formula seems  more complex to be interpreted than the relative risk one. Furthermore, the log-mean regression approach preserves the  interpretation of the decomposition of effects even for the multivariate extension.

Several works explored collapsibility conditions such that the deviation term is null and the distortion of effects can be ignored;   for instance \citet{guo-geng:1995}  derived collapsibility conditions for logistic regression models, \citet{xie-al:2008} for discrete measure of associations and \citet{didelez-al:2010} studied collapsibility conditions for odds ratio in case of outcome-dependent sampling. 

The interpretation of model parameters, and in particular of the deviation term, represent a crucial issue when the interest is  focused in modelling  rather than in exploring conditions to avoid the distortion. This generally happens in contexts where the intermediate variable plays a key role and  the  deviation term represents a relevant parameter which needs to be specifically addressed.  From this side, the approach discussed so far and, in particular, the Relative Risk formula may provide useful insights.

\section*{Acknowledgments}
I am grateful to Fabio Picciafuochi (Azienda USL, Reggio Emilia, Italy) for providing
me the data on the morphine study. I also gratefully acknowledge useful discussions with Giovanni M. Marchetti and Nanny Wermuth. 

\section*{Appendix}\label{sec:app}
\textit{Proof of Proposition \ref{thr.main}}
\begin{proof}
Consider the log-mean regression model for $Y|\{Z\cup X\}$:
	\begin{equation}\label{eq:thr1}
		\log \pi_{Y|i_{ZX}}= \alpha_{Y|ZX}+ \theta_{Y|Z.X}Z+\theta_{Y|X.Z}X+\theta_{Y|ZX}ZX	, \quad i_{ZX} \in \{0,1\}^2.
	\end{equation}
The marginal log-mean regression for $Y|X$ is obtained by summing in Equation \eqref{eq:thr1} for both levels of $Z$ in $\{0,1\}$, then,
	\begin{eqnarray*}
		\pi_{Y|\{X=1\}}&=& \exp(\alpha_{Y|ZX}+ \theta_{Y|X.Z}) \times [\exp(\theta_{Y|Z.X}+ \theta_{Y|ZX}) \pi_{Z|\{X=1\}} + (1-\pi_{Z|\{X=1\}})],
	\end{eqnarray*}	
	and
	\begin{eqnarray*}
		\pi_{Y|\{X=0\}}&=& \exp(\alpha_{Y|ZX}) \times [\exp(\theta_{Y|Z.X}) \pi_{Z|\{X=0\}} + (1-\pi_{Z|\{X=0\}})].
	\end{eqnarray*}		
Then, the result follows because $\theta_{Y|X}=\log(\pi_{Y|\{X=1\}})-\log(\pi_{Y|\{X=0\}})$ and 
$$
\theta_{Y|X}= \theta_{Y|X.Z} + \log \frac{\exp(\theta_{Y|Z.X}+ \theta_{Y|ZX})* \exp(\alpha_{Z|X}+\theta_{Z|X}) + 1-\exp(\alpha_{Z|X}+\theta_{Z|X})}{\exp(\theta_{Y|Z.X})* \exp(\alpha_{Z|X}) + 1-\exp(\alpha_{Z|X})},
$$ 	
with $\pi_{Z|\{X=1\}}=\exp(\alpha_{Z|X}+\theta_{Z|X})$ and $\pi_{Z|\{X=0\}}=\exp(\alpha_{Z|X})$.
	\end{proof}

\textit{Proof of Proposition \ref{thr.mult}}
\begin{proof}
For $|D|=1$ the result follows from Theorem \ref{thr.main}. For every non-empty subset $D \subseteq V$ with $|D|>1$, the conditional product outcome $Y^D|\{Z,X\}$ is a Bernoulli distribution with probability parameter $\mu_{D|i_{ZX}}$. Then, the result is obtained by applying the  proof of Theorem \ref{thr.main} to Equation \eqref{eq:mult1}, for any $D \subseteq V$.
\end{proof}

\textit{Proof of Theorem \ref{thr.multVU}}
\begin{proof}
Given the  regression model in Equation \eqref{eq:regmultVU}, consider the marginal model 
obtained marginalizing over $Z_U$. Then,
\begin{eqnarray*}
		\pi_{D|\{X=1\}}&=& \exp(\alpha_{D|UX}+ \theta_{D|X.U}) \times \\
		&\times& \Bigg \{\sum_{E \subseteq U}\exp \Bigg [\sum_{u \subseteq E}\theta_{D|u.X}Z_{u} \Bigg ] P(Z_E=1_E,Z_{U\setminus E}=0_{U \setminus E}|X=1) \Bigg\} ,
	\end{eqnarray*}	
	and
	\begin{eqnarray*}
		\pi_{D|\{X=0\}}&=& \exp(\alpha_{D|UX}) \times \Bigg \{\sum_{E \subseteq U}\exp \Bigg [\sum_{u \subseteq E}\theta_{D|u.X}Z_{u}\Bigg ] P(Z_E=1_E,Z_{U\setminus E}=0_{U \setminus E}|X=0) \Bigg\}.
	\end{eqnarray*}
Then, the result follows because $\theta_{D|X}=\log(\pi_{D|\{X=1\}})-\log(\pi_{D|\{X=0\}})$. 
\end{proof}

\bibliographystyle{chicago}
\bibliography{gamma-ref}
\end{document}